# A lateral resolution metric for static single molecule localization microscopy images from time-resolved pair correlation functions

**Thomas R. Shaw[1,2], Frank J. Fazekas[1], Sumin Kim[3], Jennifer C. Flanagan-Natoli[1], Emily R. Sumrall[1], S. L. Veatch[1,2]**

[1]Program in Biophysics, [2]Program in Applied Physics, [3]Program in Cellular and Molecular Biology, University of Michigan, Ann Arbor, Michigan

**ABSTRACT**

Single molecule localization microscopy (SMLM) permits the visualization of cellular structures an order of magnitude smaller than the diffraction limit of visible light, and an accurate, objective evaluation of the resolution of an SMLM dataset is an essential aspect of the image processing and analysis pipeline. Here we present a simple method that uses the pair autocorrelation function evaluated both in space and time to measure the time-interval dependent point-spread function of SMLM images of static samples.  Using this approach, we demonstrate that experimentally obtained images typically have effective point spread functions that are broader than expected from the localization precision alone, due to additional uncertainty arising from drift and drift correction algorithms.  This resolution metric reports on how precisely one can measure pairwise distances between labeled objects and is complementary to the commonly used Fourier Ring Correlation metric that also considers spatial sampling. The method is demonstrated on simulated localizations, DNA origami rulers, and cellular structures labelled by dye-conjugated antibodies or fluorescent fusion proteins.

## INTRODUCTION

Localization microscopy is a powerful tool to image structures in cells with dimensions ranging between tens of nanometers to tens of microns. Methods such as (d)STORM (1, 2), (F)PALM (3, 4), and PAINT (5) exploit the stochastic blinking of single fluorophores to localize emitting molecules with a localization precision much smaller than the diffraction limit of visible light, by imaging only a small subset of probes in any given image frame.  These samples are then imaged over time, and acquired localizations are typically assembled into a single reconstructed super-resolved image.

Assessing the quality of reconstructed images can be challenging as numerous factors can contribute. These factors can include the labeling density, the brightness and blinking dynamics of the fluorophore, motions of the stage or labeled molecules during acquisition, and the analytical methods used in post-processing.  One important measure of the quality of a measurement is the localization precision of single fluorophores, which is influenced many of the factors listed above.  Most localization algorithms directly return the localization precision of single fits and similar information can be extracted directly from the localizations themselves through the use of pair-correlation functions or nearest neighbor analyses that extract the distribution of positions of molecules detected in adjacent frames (6, 7). Other metrics of image quality have been developed that integrate both precision and spatial sampling, reporting superior resolution when localizations effectively sample labeled objects (8, 9).  One widely used method is called Fourier Ring Correlation (FRC) (10–13) and is analogous to resolution metrics used in x-ray crystallography.  FRC has the advantage of capturing the impact of factors that degrade quality over the entire span of image acquisition, such as sample drift and imperfect drift correction.  A disadvantage of FRC on reconstructed super-resolution images is that the absolute number produced depends on the types of structures imaged and the specific regions of interest used.

Here we present a simple method to evaluate the effective localization precision of a localization dataset that incorporates all errors relevant to measuring distances between labeled objects. This resolution measure is expected to be the most useful when the goal of the imaging experiment is to measure distances between labeled components, for example when constraining the structure of a multi-protein complex or when tabulating the local density of labeled components. The method presented exploits temporal correlations in the blinking dynamics of single fluorophores commonly used for localization microscopy (14–16). This enables the method to report on errors accumulated over time that do not typically impact the localization precision but impact how accurately a molecule's position is determined relative to others. Here, we derive the proposed resolution metric, validate it through simulation and an image of DNA origami rulers (17), and apply it to several images of labeled structures in cells.

**MATERIALS AND METHODS**

**Simulations**
Simulations mimicking DNA origami rods were accomplished by randomly placing pairs of fluorophores positioned 50nm apart within a 40μm by 40μm region of interest at an average density of 1 pair per μm$^2$. 20,000 individual image frames with an effective frame time of 0.1 sec were simulated by sampling a subset of molecular positions with a localization precision of 10nm in each lateral dimension. The dynamics of individual fluorophores were governed by a continuous time Markov process involving five states: one on state (1), three dark states (0, 01, 02), and an irreversible bleached state (B), following the procedure described previously (16, 18). The on state was accessible from any of the dark states, while dark state 0 was accessible only from the on state, and dark states 01 and 02 were accessible only from the previous dark states, 0 and 01, respectively. We used the following parameters (using the notation described in (16, 18)) to capture essential elements of our experimental observations: $\lambda(0 \rightarrow 1) = 1.2$ Hz; $\lambda(0 \rightarrow 01) = 0.05$ Hz; $\lambda(01 \rightarrow 02) = 0.0033$ Hz; $\lambda(01 \rightarrow 1) = 0.02$ Hz; $\lambda(02 \rightarrow 1) = 0.0005$ Hz; $\lambda(1 \rightarrow 0) = 5$ Hz; $\mu(1 \rightarrow B) = 0.05$ Hz. The continuous time Markov process was simulated with the MATLAB File Exchange code "simCTMC.m" (19). When present, drift was applied to all molecular positions with a constant rate of 0.3nm/sec in the x direction along with diffusive drift characterized by a diffusion coefficient of D=2.5nm$^2$/sec. Drift was corrected using the mean-shift algorithm described previously (20) using 1000 frames per alignment (10s).

**Experimental sample preparation:**
DNA origami "gatta-STORM" nanorulers were purchased from Gattaquant GMBH (Grafelfing, Germany) and a sample was prepared following manufacturer's instructions. Briefly, biotinylated Bovine Serum Albumin (BSA; ThermoFisher; 1 mg/ml) was absorbed to a clean 35mm #1.5 glass-bottom dish (MatTek well; MatTek Life Sciences) for 5min then washed. Streptavidin was then applied (1 mg/ml) for 5min, then washed with a solution of phosphate buffered saline (PBS) plus 10mM MgCl$_2$. A solution containing the biotinylated DNA origami was then applied. Samples were then washed and imaged in an imaging buffer supplemented with 10mM MgCl$_2$. "gatta-PAINT" 40R nanorulers were purchased from Gattaquant GMBH and imaged following the manufacturers recommendations.

Mouse primary neurons were isolated from P0 mouse pups as described previously and cultured on MatTek wells (20). On day 10 of culture (days *in vitro* 10), neurons were rinsed with sterile Hank's

Balanced Salt Solution and fixed for 10min with pre-warmed 4% PFA (Electron Microscopy Sciences) in Phosphate Buffered Saline (PBS). The fixed neurons were rinsed three times with PBS and permeabilized in 0.2% Triton X-100 (Millipore Sigma) in PBS for 5min. Neurons were then incubated in blocking buffer containing 5% BSA for 30min, and labeled with Nup210 polyclonal antibody diluted in PBS (1:200; Bethyl laboratories A301-795A) overnight in 4 °C. The following day, neurons were washed three times in PBS and stained with goat-anti-rabbit Alexa Fluor 647 Fab Fragment (1:800; Jackson ImmunoResearch 111-607-003) for an hour, washed three times with PBS, then imaged.

CH27 B-cells (mouse, Millipore Cat# SCC115, RRID:CVCL_7178), a lymphoma-derived cell line (21) were acquired from Neetu Gupta (Cleveland Clinic). CH27 Cells were maintained in culture as previously described (22). Cells were adhered to MatTek wells coated with VCAM following procedures described previously (23). Briefly 0.1 mg/ml IgG, Fcγ-specific was absorbed to a plasma cleaned well for 30 min at room temperature. Wells were rinsed with PBS, then nonspecific binding was blocked with 2% BSA at room temperature for 10 minutes, followed by incubation with 0.01 mg/mL recombinant human VCAM-1/CD106 Fc chimera protein (R&D Systems) and 0.01 mg/mL ChromPure Human IgG, Fc fragment (Jackson Immunoresearch) for 1 hour at room temperature or overnight at 4°C. VCAM-1 coated dishes were stored up to 1 week in VCAM-1 and Fc at 4°C. Immediately prior to plating, dishes were blocked at room temperature in 2% goat serum (Gibco) for 10 min, then cells were allowed to adhere for 15 min in media prior to chemical fixation in 2% PFA and 0.2% glutaraldehyde (Electron Microscopy Sciences). F-Actin was stained by permeabilizing cells with 0.1% Triton-X-100 prior to incubation with 3.3 µM phalloidin-Alexa647 (Invitrogen) for at least 15 min. Phalloidin stained cells were imaged immediately after removing label. Cells transiently expressing Clathrin-GFP were permeabilized after fixation with 0.1% Triton-X-100 followed by labeling with a single domain anti GFP antibody (MASSIVE-SDAB 1-PLEX) from Massive Photomics GMBH (Grafelfing, Germany) for 1h at room temperature, then imaged in 0.5nM of imaging strand in the imaging buffer supplied by the manufacturer.

Cells expressing the membrane label Src15-mEos3.2 were prepared by transiently transfecting $10^6$ cells with a 1 µg of plasmid encoding Src15-mEos3.2 (N'-MGSSKSKPKDPSQRRNNNNGPVAT-[mEos3.2]-C') which was derived from a GFP tagged version (24). Transfection was accomplished by Lonza Nucleofector electroporation (Lonza, Basel, Switzerland) with program CA-137 and cells were grown in flasks overnight prior to plating and fixation as described above.

**Single molecule imaging and localization**
Imaging was performed using an Olympus IX83-XDC inverted microscope. TIRF laser angles where achieved using a 100X UAPO TIRF objective (NA = 1.49), and active Z-drift correction (ZDC) (Olympus America). Alexa 647 was excited using a 647 nm solid state laser (OBIS, 150 mW, Coherent) and mEos3.2 was excited using a 561nm solid state laser (Sapphire 561 LP, Coherent), both coupled in free-space through the back aperture of the microscope. Fluorescence emission was detected on an EMCCD camera (Ultra 897, Andor). Samples containing Alexa647 were imaged in a buffer containing 100mM Tris, 10mM NaCl, 550mM glucose, 1% (v/v) beta-mercaptoethanol, 500 µg/ml glucose oxidase (Sigma) and 40 µg/ml catalase (Sigma), with 10mM $MgCl_2$ for the DNA origami sample. Samples with mEos3.2 or DNA PAINT probes were imaged in imaging buffer from Massive Photomics GMBH. Single molecule positions were localized in individual image frames using custom software written in Matlab. Peaks were segmented using a standard wavelet algorithm (25) and segmented peaks were then fit as single emitters on GPUs using previously described algorithms for 2D (26), or as multi-emitters on a CPU using the ThunderStorm ImageJ plugin (27). After localization, points were culled to remove outliers prior to drift correction (20). Images were rendered by generating 2D histograms from localizations followed by convolution with a Gaussian for display purposes. For the nano-ruler samples, localizations were

$\epsilon$ = 12 nm and minPts = 15.

**Evaluation of space-time autocorrelations.**
Space-time autocorrelations were tabulated by first tabulating space- and time- displacements between all pairs of localizations within a specified region of interest (ROI) detected in a given dataset. This was accomplished using a *crosspairs()* function based on the one from the R package spatstat (1), but used here as a C routine with a MATLAB interface, as described previously(20). Lists of displacements were converted into space-time autocorrelation functions by binning in both time and space within the C routine for improved performance, followed by a normalization implemented in Matlab that produces a value g(r, \tau)=1 when localizations are randomly distributed in both space and time within the specified ROI. A derivation of the form of this normalization and an explanation of how it is computed is presented in Supplementary Note 1.

**Estimation of $g_{PSF}(r,\tau)$ and $\sigma_{xy}(\tau)$**

The core computations of the resolution estimation are gathered in a single MATLAB function. First, $g(r_i,\tau_j)$ is computed as described above, for a range of distance and time separation values. By default, $r_i, i=1,\ldots,N_r$ range from 2.5 to 250nm, with equal spacing of 5 nm, and $\tau_j, j=1,\ldots,N_\tau-1$ are log-spaced, with the final time separation $\tau_{max}=\tau_{N_\tau}$ given by three quarters of the length of the dataset in time. Both the $r_i$ and the $t_j$ may optionally be specified by the user to override these defaults. Then, $\Delta g(r,\tau_j) = g(r,\tau_j) - g(r,\tau_{max})$ are computed for each $\tau$, normalized by their first points, and fitted to a Gaussian of the form $A*\exp(-r^2/4\sigma_{xy,j}^2)$, using MATLAB's nonlinear least squares fitting routing fit(). $\sigma_{xy,j}$ is reported as the estimate of $\sigma_{xy}(\tau_j)$. Bootstrapped standard errors are determined by choosing 8 subsamples of the points, each containing ¼ as many points as the full dataset, and estimating $\sigma_{xy}(\tau_j)_k$ for each subsample $k$, in the same way as for the full dataset. The standard error is reported as .5*std. dev. $(\sigma_{xy}(\tau_j)_k$, where the .5 accounts for the overestimate of errors due to using 4 times fewer points.

**Measuring $g_{PSF}(r,\tau)$ in simulations by grouping localizations with molecules**

In simulations, localizations imaged at $x_i, y_i, \tau_i$ are associated with the molecules that produced them. We tabulate displacements between all pairs associated with the same molecule $\Delta r_{ij}=\sqrt{\Delta x_{ij}+\Delta y_{ij}}$ and $\Delta t_{ij}$. The list of all pairs is binned into two dimensional histograms following the same r and τ bin-edges as described for computing $g(r,\tau)$ above and are normalized by the number of pairs contributing to each bin. Distributions at each τ bin are fit to the same Gaussian form as applied to the $g_{PSF}(r,\tau)$ estimated from $\Delta g(r,\tau)$.

**Determining the resolution with Fourier Ring Correlation**

The resolution of each dataset was assessed with Fourier Ring Correlation (FRC) (12). To produce the FRC curves, localizations were divided into consecutive blocks of 500 frames, and these blocks were randomly placed into one of two statistically independent subsets. For the simulated and experimental DNA origami datasets, as well as the nuclear pore complex dataset, the pixel size for the FRC calculation was taken to be 5nm, and square regions 10μm on a side were used as a mask. For the actin and src15 datasets, the pixel size was 10nm, with the mask 20μm on each side. 20 randomly determined repetitions of the calculation were performed for each dataset.

**RESULTS**

**Derivation of the resolution metric.**

The spatial autocorrelation function describing a distribution of static molecules is given by $g_{molecules}(r)$ and is tabulated as described in Methods and Supplementary Materials. This function can be divided into two components:

$$g_{molecules}(\vec{r}) = \frac{1}{\rho}\delta(\vec{r}) + g_p(\vec{r}). \tag{1}$$

The first term in Eqn. 1 comes from counting single emitters and is a delta function ($\delta(\vec{r})$) with magnitude equal to the inverse average density of molecules ($\rho$) over the region of interest (ROI). The second term in Eqn. 1 comes from correlations between distinct pairs of molecules and reports on the detailed structure present in the image. In SMLM, single emitters labeling molecules have dynamics governed by the probe photo-physics, which can be described with the temporal autocorrelation function $g_e(\tau)$. Probes can remain on for multiple sequential image frames and can blink on again at a later time before eventually bleaching irreversibly (14–16). As a result, $g_e(\tau)$ is highly correlated (>1) at short time-intervals and decays sharply on time-scales describing the average on-time of fluorophores. This function continues to decay slowly at long τ, both because some probes tend to flicker over medium to long time-scales and because some fluorophores eventually bleach. Including $g_e(\tau)$ produces the following spatio-temporal autocorrelation function for the emitting molecules:

$$g_{emitters}(\vec{r},\tau) = \frac{1}{\rho}\delta(\vec{r})g_e(\tau) + g_p(\vec{r}). \tag{2}$$

Eqn. 2 assumes the blinking statistics of different fluorophores are uncorrelated, which is why $g_e(\tau)$ multiplies only the first term.

When fluorophores are observed with finite spatial resolution, the observed autocorrelation is blurred (convolved) by the autocorrelation function that describes that resolution, which in microscopy is typically called the point spread function or PSF, or $g_{PSF}(\vec{r},\tau)$. Including this factor, the autocorrelation function for the observed image, $g(\vec{r},\tau)$, becomes:

$$g(\vec{r},\tau) = \frac{1}{\rho}g_{PSF}(\vec{r},\tau)g_e(\tau) + g_{PSF}(\vec{r},\tau) * g_p(\vec{r}) \tag{3}$$

where * indicates a convolution. The first term in Eqn. 3 describes multiple observations of the same molecule and contains the most information on the PSF of the measurement. To isolate this component, $\tau = \tau_{max}$ where the value of $g_e(\tau)$ is relatively small.

$$\Delta g(\vec{r},\tau) = g(\vec{r},\tau) - g(\vec{r},\tau_{max})$$
$$\approx \frac{1}{\rho}\left(g_{PSF}(\vec{r},\tau)g_e(\tau) - g_{PSF}(\vec{r},\tau_{max})g_e(\tau_{max})\right)$$
$$\approx \frac{1}{\rho}g_{PSF}(\vec{r},\tau)\left(g_e(\tau) - g_e(\tau_{max})\right)$$

Where an appropriate choice of $\tau_{max} \gg \tau$ depends on the length and blinking dynamics that are present in a given dataset. More simply written:

$$g_{PSF}(\vec{r},\tau) \approx \frac{\rho}{\Delta g_e(\tau)}\Delta g(\vec{r},\tau) \qquad (4)$$

In words, Eqn. 4 states that the shape of the point spread function can be estimated by comparing autocorrelation functions tabulated at different time-intervals. This observation forms the basis for the method demonstrated in this report.

The approximation of Eqn. 4 is valid when $g_{PSF}(\vec{r},\tau)$ does not depend strongly on $\tau$. This allows for the cancelation of the term involving $g_p(\vec{r})$ which captures image-specific factors that otherwise complicate the measurement of $g_{PSF}(\vec{r},\tau)$. Moreover, Eqn 4. is only valid in the limit where $g_{PSF}(\vec{r},\tau)g_e(\tau) - g_{PSF}(\vec{r},\tau_{max})g_e(\tau_{max}) \approx g_{PSF}(\vec{r},\tau)\Delta g_e(\tau)$. This assumption is expected to be valid for short and medium time-intervals when $g_e(\tau) \gg g_e(\tau_{max})$, or when $g_{PSF}(\vec{r},\tau) \approx g_{PSF}(\vec{r},\tau_{max})$. Under both of these conditions, subtraction of $g_{PSF}(\vec{r},\tau_{max})$ does not meaningfully distort the shape of $g_{PSF}(\vec{r},\tau)$, as demonstrated in Supplementary Figure S1. The assumption that $g_{PSF}(\vec{r},\tau) \approx g_{PSF}(\vec{r},\tau_{max})$ is likely justified in most practical cases for larger τ, since drift correction algorithms are most effective at correcting errors at slow time-scales. For cases where $g_{PSF}(\vec{r},\tau)$ broadens at long time-intervals, for example when labeled molecules diffuse over length-scales comparable to the localization precision over the acquisition time, we expect that Eqn. 4 will produce a systematically narrow estimate of $g_{PSF}(\vec{r},\tau)$ for longer τ when $g_e(\tau) \sim g_e(\tau_{max})$, as demonstrated on simulated data in Supplementary Figure S2. In principle, more sophisticated analytical methods could be applied to independently extract $g_{PSF}(\vec{r},\tau)$ and $g_{PSF}(\vec{r},\tau_{max})$ for cases where $g_e(\tau) \sim g_e(\tau_{max})$. Alternately, a user could tabulate $\Delta g(\vec{r},\tau)$ for closely spaced τ where changes in $g_{PSF}(\vec{r},\tau)$ are expected to be more subtle. For the purposes of this report, we indicate when estimates of $g_{PSF}(\vec{r},\tau)$ may be subject to this systematic bias, using a cutoff of $\Delta g(r<5nm,\tau)/g(r<5nm,\tau_{max}) < 0.5$.

The applicability of Eqn 4. Is dependent on the form of $g_e(\tau)$

$g(r<25nm,\tau)$ capture the shape of $g_e(\tau)$ up to a numerical offset, and several examples showing this decay for Alexa647 and mEos3.2 fluorophores are shown in Supplementary Figure S3. The slow decay of $g_e(\tau)$ at longer time-intervals means that the statistical power of $\Delta g(\vec{r},\tau)$ will degrade since fewer correlated pairs are observed at these long-time intervals. To increase statistical significance, we group time-intervals into increasingly large windows, typically log-spaced in τ to account for the exponential decay inherent in $g_e(\tau)$. For the examples shown in this report, $g(\vec{r},\tau_{max})$ is evaluated over $\frac{1}{2}T_{max} < \tau_{max} < T_{max}$, where $T_{max}$ is the total acquisition time. Statistical confidence is estimated through bootstrapping and we estimate the statistical power of $\Delta g(\vec{r},\tau)$ directly from $g(\vec{r},\tau)$, as described in Materials and Methods.

To convert $g_{PSF}(\vec{r},\tau)$ to a resolution length-scale, we assume that it takes on a Gaussian form:

$$g_{PSF}(\vec{r},\tau) \propto \exp\{-x^2/4\sigma_x^2(\tau) - y^2/4\sigma_y^2(\tau)\}, \tag{5}$$

where $\sigma_x^2(\tau)$ is the standard deviation in the $x$ direction of the distance between the true position of the molecule at time $t$ and the and a localization at time $t+\tau$. The extra factor of 2 in the denominator accounts for the fact that $g_{PSF}$ reports on the distribution of distances between pairs of localizations, resulting in twice the variance compared to the error in one localization. Typically image resolution is isotropic in the lateral dimensions so we take $\sigma_{xy} := \sigma_x = \sigma_y$ and compute angularly averaged correlation functions resulting in:

$$g_{PSF}(r,\tau) \propto \exp\left\{-\frac{r^2}{4\sigma_{xy}(\tau)^2}\right\} \tag{6}$$

It is convenient to also define the mean squared displacement $\sigma_r^2(\tau) = \sigma_x^2(\tau) + \sigma_y^2(\tau) = 2\sigma_{xy}^2(\tau)$ to evaluate localization precision, which accounts for errors in both dimensions. When localizations are acquired in three dimensions, the axial resolution often differs and this component can be considered independently:

$$g_{PSF}(z,\tau) \propto \exp\left\{-\frac{z^2}{4\sigma_z(\tau)^2}\right\}$$

**Validation through Simulation**

To validate this resolution metric, we generated simulated datasets of DNA origami nanorulers in which fluorophores are separated by a fixed distance of 50 nm. The blinking of the fluorophores were subject to a photophysical model based on (16, 18). Briefly, fluorophores could exist in an "on" state, one of three dark states, or a bleached state. Transitions between states were governed by a continuous time Markov process, with transition rates roughly based on those measured in (16) but modified to reflect the experimental conditions used to obtain experimental images in this work. Nanorulers were placed randomly and uniformly with an average density of 1/μm² across a 40μm by 40μm field of view with the molecules having a localization precision of 10 nm in each lateral dimension. 20,000 image frames were simulated with a frame-time of 0.1 s. Figure 1a illustrates a small field of view containing 3 nanorulers, both as a reconstructed image and with localizations colored by time. An image showing a larger subset of the field of view is shown as Supplementary Figure S4.

Simulated localizations were subjected to a spatiotemporal auto-correlation analysis as described in Methods and representative plots of the spatial component of $g(r,\tau)$ are shown in Fig 1b. This family of curves contains two major features: an initial peak at short displacements (r<40nm) arising from multiple localizations from the same molecule, and a second feature at wider radii (40nm<r<100nm) arising from displacements between localizations from different molecules on the same ruler. The amplitude of the initial peak decreases with increasing τ, while the second feature is largely independent of τ. The τ dependent component is isolated by subtracting $g(r,\tau)$ at long τ from those arising from shorter τ to obtain $\Delta g(r,\tau)$ as shown in Fig 1d. In this simulation, there are no τ dependent effects that would impact resolution, resulting in $\Delta g(r,\tau)$ having the same width for all τ. This is summarized by fitting $\Delta g(r,\tau)$ to the Gaussian function of Eqn. 6 to extract $\sigma_{xy}$ which is reported in Fig 1e. In this simulated example, we can associate all localizations with the molecules that produced them, and can therefore directly compute $g_{PSF}(r,\tau)$ from localizations as described in Methods and the resulting resolution obtained is also plotted in Fig 1e. Lastly, we tabulate displacements between all localizations originating from distinct molecules on the same ruler. The distribution is shown in Fig 1(f) and its properties are described by simulation parameters. The line in Fig 1f has a Gaussian shape with the form: $P(r) \propto \exp\{-(r-\langle r \rangle)^2/4\sigma_{xy}^2\}$ where $\sigma_{xy}$ is 10nm and $\langle r \rangle = \sqrt{50^2 + 2\sigma_{xy}^2}$ is 51.96nm. The slight bias in $\langle r \rangle$ towards a value larger than separation between molecules (50nm) arises from the components of localizations that fall perpendicular to the ruler axis and always contribute positive values to the measured displacements. For comparison, the FRC was also tabulated to be 30nm for this simulation following procedures described in Methods. This FRC value is slightly larger than $2\sigma_r = 2\sqrt{2}\sigma_{xy} = 28$nm.

The simulation of Fig 1 does not contain any factors expected to degrade image resolution over time. In Figure 2, the same simulation is subjected to a directed drift in the x-direction as well as diffusive drift in both x and y, then drift is corrected using a mean shift algorithm (20) that works by evenly dividing localizations into time-bins, then finding the displacement that minimizes the mean distance between localizations across all time-bins. The applied drift and the calculated drift correction are shown in Fig 2a along with the resulting image reconstruction. $g(r,\tau)$

$$\Delta g(r,\tau)$$

$$\langle \sigma_{xy} \rangle = 11.3 nm,$$

which is determined by averaging over estimated $\sigma_{xy}(\tau)$ weighted by the number of pairs associated with each time-interval window. The FRC resolution of the drift-corrected dataset in Figure 2 is 35 nm. This again is somewhat larger than $2\langle \sigma_r \rangle = 32$ nm. Observing a plateau in plots of $\sigma_{xy}(\tau)$ is a good indicator that the algorithm is successfully applied, since drift correction is designed to stabilize errors on long time-scales. Supplementary Figure S2 shows an example of the same simulation with drift and drift correction, but where individual molecules are allowed to diffuse slowly such that $g_{PSF}(r)$ increase with τ in a way that is not accounted for through drift correction. In that case, $\sigma_{xy}(\tau)$ increases with τ and is underestimated by $\Delta g(r,\tau)$.

**Estimating the resolution of experimental SMLM datasets**
Figure 3 demonstrates this approach on an experimental dataset of DNA origami nanorulers that resemble the simulated rulers with Alexa647 labeling sites separated by 50nm. Fig 3a shows a small subset of the field of view of the acquired image, that was reconstructed from 29,000 image frames acquired over 53 min at a frame rate of 0.1s, with a total of over 126,000 individual localizations. In post-processing, a drift correction was applied with a time-window width of 25s or 250 image frames. As in the simulated case, $g(r,\tau)$ decays at short r with increasing $\tau$ (Fig 3c), and $\Delta g(r,\tau)$ is roughly Gaussian (Fig 3d). Fitting $\Delta g(r,\tau)$ yields the resolution $\sigma_{xy}(\tau)$. As in the simulated example, the measured resolution is lowest at short time-intervals (close to 7nm) and plateaus at time-scales somewhat shorter than the frequency of the applied drift correction.

Since the localization clouds from individual Alexa647 molecules were visually distinct, we applied a DBSCAN segmentation algorithm to associate localizations with individual molecules. From this segmentation, we tabulated the pairwise distances between molecules on the same origami and the distribution of these values is shown in Fig 3e. This distribution is well described by a model applying the measured $\langle \sigma_{xy}(\tau) \rangle = 7.7$ nm with $\langle r \rangle = 52.2 \pm 0.2$ nm, where the error is dominated by uncertainty in the sample magnification at the camera. This yields a separation distance of $51.0 \pm 0.2$ nm between origami on individual rulers, which is within the manufacturer's specifications. The FRC resolution was found to be 19 nm, this time slightly smaller than $2\langle \sigma_r \rangle = 21.8$ nm.

For demonstration, we have conducted this same analysis on a similar DNA origami sample that was imaged using DNA PAINT, this time using rulers containing 3 docking sites separated by 40nm and summarized in Supplementary Figure S6. In contrast to the dSTORM fluorophores of Fig 3, molecules imaged by DNA PAINT do not exhibit long time-scale correlations, limiting the applicability of this method. The DNA PAINT probes used for this image do remain correlated over time-scales relevant for drift-correction (>15s), which in itself provides a useful estimate of image resolution.

We next apply this method to image labeled structures in cells. Figure 4 shows the method applied to nuclear pore complexes within the nuclear envelope of chemically fixed primary mouse neurons. In these images, a protein component of NPCs, NUP210, was labeled with a conventional primary antibody and a fAb secondary directly conjugated to Alexa647. 12500 images were acquired over 23 min with an integration time of 0.1s and a total of 178873 localizations detected on the nuclear envelope. Drift correction was accomplished with a time-window of 12.5 s or 125 image frames. Reconstructed images of the entire nucleus and single pores are shown in Fig 4a. along with a scatter plot demonstrating that individual NPC subunits are sampled at times throughout the observation. $g(r,\tau)$

$\Delta g(r,\tau)$ (Fig 4c.). Fitting $\Delta g(r,\tau)$ to a Gaussian shape

$\tau > 30s$ ), $\Delta g(r < 5nm, \tau)/g(r < 5nm, \tau_{max})$ falls below the value of 0.5, indicating that the estimate of $g_{PSF}(r, \tau)$ from $\Delta g(r, \tau)$ may be subject to distortion from the subtraction of $g(r, \tau_{max})$. In this case, we expect the estimate remains valid, both because this threshold is crossed beyond the drift-correction time-scale and because $\sigma_{xy}(\tau)$ remains constant in this range. Including these points, we estimate $\langle \sigma_{xy} \rangle$ to be 9.2nm. The FRC resolution was evaluated to be 37 nm, which is larger than $2\langle \sigma_r \rangle = 26$ nm.

Supplemental Figure S7 shows a similar class of cellular structure imaged using DNA PAINT. In this example, clathrin-gfp is transiently expressed in CH27 cells then labeled post fixation with an anti GFP nanobody conjugated to an ssDNA docking strand. Cells are then imaged in the presence of a complementary imaging strand labeled with Atto 655. Similar to the origami DNA PAINT sample of Supplemental Figure S6, temporal correlations from single molecules remain for short to medium time-scales (<10s), allowing for accurate estimation of resolution losses due to drift and drift correction. Beyond 10s, the estimate of $g_{PSF}(r, \tau)$ from $\Delta g(r, \tau)$ fails, as indicated by $\Delta g(r < 5nm, \tau)/g(r < 5nm, \tau_{max})$ falling well below 0.5 and a systematic upward trend in $\sigma_{xy}(\tau)$.

Figure 5 shows the method applied to an image of f-actin staining by phalloidin-Alexa647 in chemically fixed CH27 B cells adhered to a glass surface decorated with VCAM. For this sample, 5000 images were acquired over 4.9 min with an integration time of 0.05s and a total of 604056 localizations. Drift correction was accomplished with a time-window of 6.3s or 125 image frames. Unlike Figs 3 and 4 where labels decorate isolated structures scattered over a surface, this reconstructed image of f-actin is more space filling, making up a web of fibers that extend across the entire ventral cell surface (Fig 5a.). This extended structure can be detected in $g(r, \tau)$ (Fig 5b.) as increased intensity in the tail that extends to large separation distances for curves generated at all τ. This large-scale structure is effectively removed in $\Delta g(r, \tau)$ (Fig 5c) allowing for a determination of image resolution over a range of time-scales as shown in Fig 5d. Again, image resolution increases with τ until the drift-correction time scale and then plateaus, giving us confidence that the calculation remains valid even though $\Delta g(r < 5nm, \tau)/g(r < 5nm, \tau_{max})$ falls below the threshold of 0.5. Including all pairs, the estimate for $\langle \sigma_{xy} \rangle$ is 13.5nm. The FRC resolution was found to be 290 nm, which is much larger than $2\langle \sigma_r \rangle = 38nm$.

As a final demosntration, Figure 6 shows the method applied to an image of Src15, a myristoylated peptide bound to the inner leaflet of the plasma membrane and directly conjugated to the photo-switchable protein fluorophore mEos3.2. This peptide uniformly decorates the ventral surface of a chemically fixed CH27 B cell adhered to a glass surface decorated with VCAM, as seen in the reconstructed image of Fig 6a. For this sample, 7000 images were acquired over 12.7 min with an integration time of 0.1s and a total of 240,503 localizations. Drift correction was accomplished with a time-window of 12.5s or 125 image frames. mEos3.2 exhibits different blinking dynamics than Alexa647, with some probes exhibiting correlated blinking on long time-scales. This can be seen in plots of $g(r, \tau)$ $\Delta g(r, \tau)$ curves isolate the initial peak, allowing for the quantification of image resolution. In this example, the slow decay of $g(r < 100nm, \tau)$ with τ allows for better time-resolution of the resolution metric, since the amplitude of $\Delta g(r, \tau)$ remains large over a broad range of timescales. The FRC resolution was found to be 57 nm, which is larger than $2\langle\sigma_r\rangle = 38$ nm.

**CONCLUSIONS:**

Here we present a resolution metric that estimates the effective average point spread function of a super-resolution fluorecence localization measurement from acquired data, relying on a few reasonable assumptions. The basic method is validated through simulations and demonstrated using experimental data of two commonly used localization microscopy probes. Importantly, the described method performs best when used alongside flourophores that exhibit blinking dynamics that remain correlated in time out to time-scales relevant to sources of error present in the imaging experiment. This resolution metric directly reports on how accurately the positions of molecules are recorded at the end of an experimental and analytical pipeline, allowing experimenters to validate and optimize new methods. Furthermore, it explicitly probes time-dependent errors that emerge due to motion of the microscope stage and/or individual fluorophores. Beyond optimization, we expect this method to be useful when interpreting experiments that involve the measurement of distances between localizations in images. It is a complimentary approach to resolution metrics that also incorperate information about spatial sampling.

The reported method estimates resolution by fitting the estimated PSF at each time-interval probed to a Gaussian shape, followed by a weighted average to extract the estimated error of the average localization in an image. This approach is convenient because the resolution is summarized as a single number, but removes information extracted from the acquired image. For example, removing fitting would yield a full average PSF that could be used for other purposes, such as deconvolution of reconstructed images or tabulated spatial correlation functions, or as input to clustering algorithms or other analyisis tools.

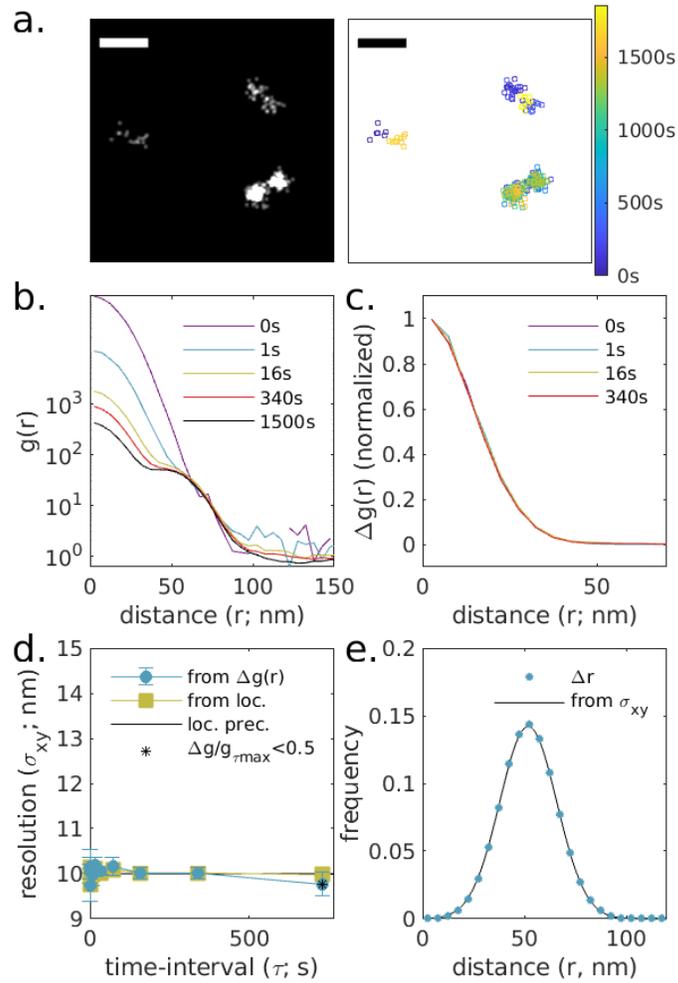

**Figure 1:** Validation of approach through simulation. (a.) Simulations consist of randomly positioned pairs of molecules positioned 50nm apart in two dimensions. Reconstructed image (left) and scatterplot of localizations with color representing the observation time (right) for a small subset of the simulated plane. Scale-bar is 100nm. A reconstructed image showing a larger field of view is shown in Supplementary Figure S4. (b.) Auto-correlations as a function of displacement, $g(r,\tau)$, tabulated from simulations for time-interval windows centered at the values shown. (c.) $\Delta g(r,\tau) = g(r,\tau) - g(r,\tau = 1500s)$ for the examples shown in b. (d.) $\Delta g(r,\tau)$ are fit to $\Delta g(r,\tau) \propto \exp\{-r^2/4\sigma_{xy}^2\}$ to extract out the resolution in each lateral dimension (from $\Delta g(r)$), which in this case is the same as the resolution deduced by grouping localizations with their associated molecules (from loc.) and the localization precision (loc. prec.) 10nm) at all time-intervals. Here, $\Delta g(r<5nm,\tau)/g(r<5nm,\tau_{max})$ falls below 0.5 only for the largest τ shown, but this is not expected to impact results since $g_{PSF}(r)$ is invariant in τ. (e.) The distribution of displacements between different molecules on the same ruler are well described by a model incorporating the localization precision (10nm) and the separation distance (50nm).

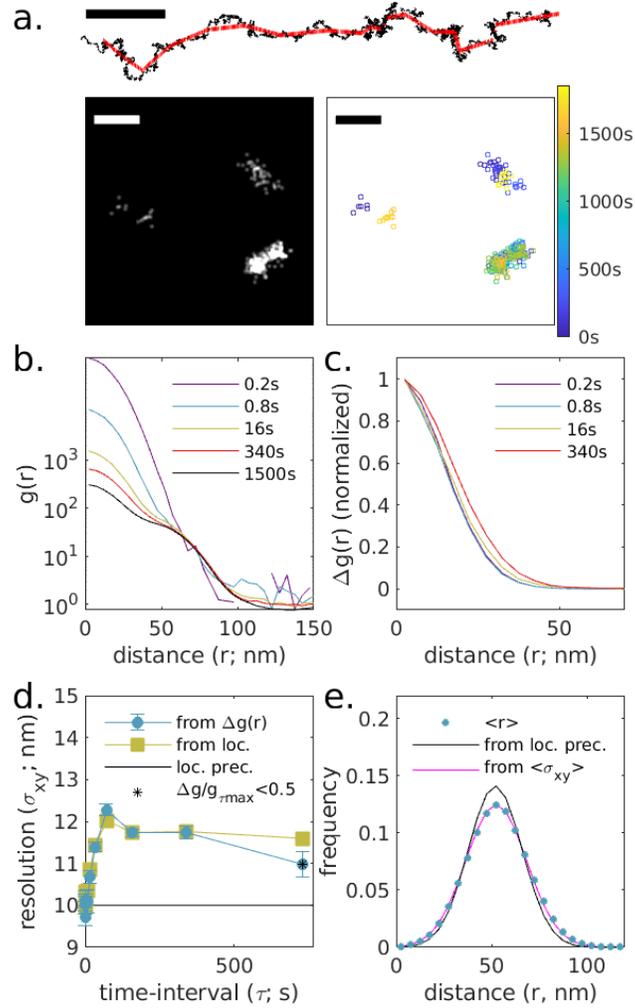

**Figure 2:** Validation of approach through simulation with drift and drift correction. (a.) The simulation from Fig 1 with applied drift (black) and drift correction (red) as shown in the trajectory above. Reconstructed image (left) and scatterplot of localizations with color representing the observation time (right) for a small subset of the simulated plane. Scale-bar is 100nm. (b.) Auto-correlations as a function of displacement, $g(r,\tau)$, tabulated from simulations for time-interval windows centered at the values shown. (c.) $\Delta g(r,\tau) = g(r,\tau) - g(r, \tau=1500s)$ for the examples shown in b. (d.) $\Delta g(r,\tau)$ are fit to $\Delta g(r,\tau) \propto \exp\{-r^2/4\sigma_{xy}^2\}$ to extract out the resolution in each lateral dimension (from $\Delta g(r)$). In this case, the resolution estimated from $\Delta g(r,\tau)$ varies with time-interval, closely following the point spread function measured by grouping localizations with molecules ((from loc.). $\Delta g(r<5nm,\tau)/g(r<5nm,\tau_{max})$ falls below 0.5 only for the largest τ shown. (e.) The distribution of displacements between different molecules on the same ruler are well described by a model incorporating the average estimated resolution ($\langle\sigma_{xy}\rangle = 11.3nm$ and the separation distance (50nm).

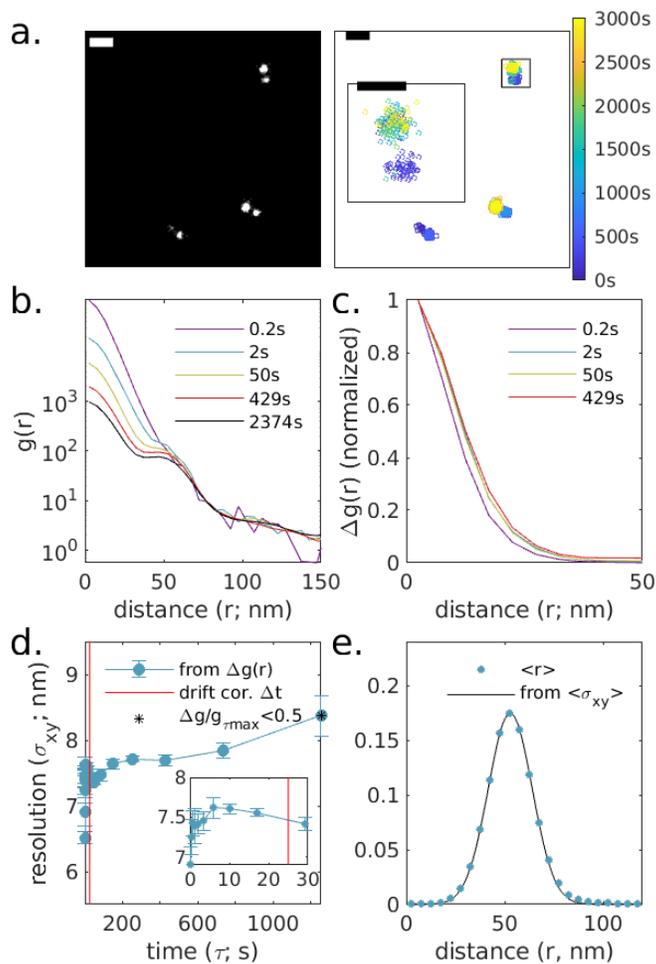

**Figure 3:** Experimental observations of DNA origami rulers. (a.) Reconstructed image (left) and scatterplot of localizations with color representing the observation time (right) for a small subset of the observed plane. Scale-bar is 100nm and 50nm in the inset. A larger field of view from this image is shown in Supplementary Figure S5. (b.) Auto-correlations as a function of displacement, $g(r,\tau)$, tabulated from localizations for time-interval windows centered at the values shown. (c.) $\Delta g(r,\tau) = g(r,\tau) - g(r,\tau = 2374s)$ for the examples shown in b. (d.) $\Delta g(r,\tau)$ are fit to $\Delta g(r,\tau) \propto \exp\{-r^2/4\sigma_{xy}^2\}$ to extract the resolution in each lateral dimension (from $\Delta g(r)$) which varies with time-interval. $\Delta g(r<5nm,\tau)/g(r<5nm,\tau_{max})$ falls below 0.5 only for the largest τ shown. The average resolution for this image is $\langle\sigma_{xy}\rangle = 7.7nm$. (e.) The distribution of displacements between different molecules on the same ruler, segmented using DBSCAN as described in Methods. Fitting to a Gaussian shape with width given by the measured $\langle\sigma_{xy}\rangle$ produces $\langle r \rangle = 52.2 \pm 0.2 nm$.

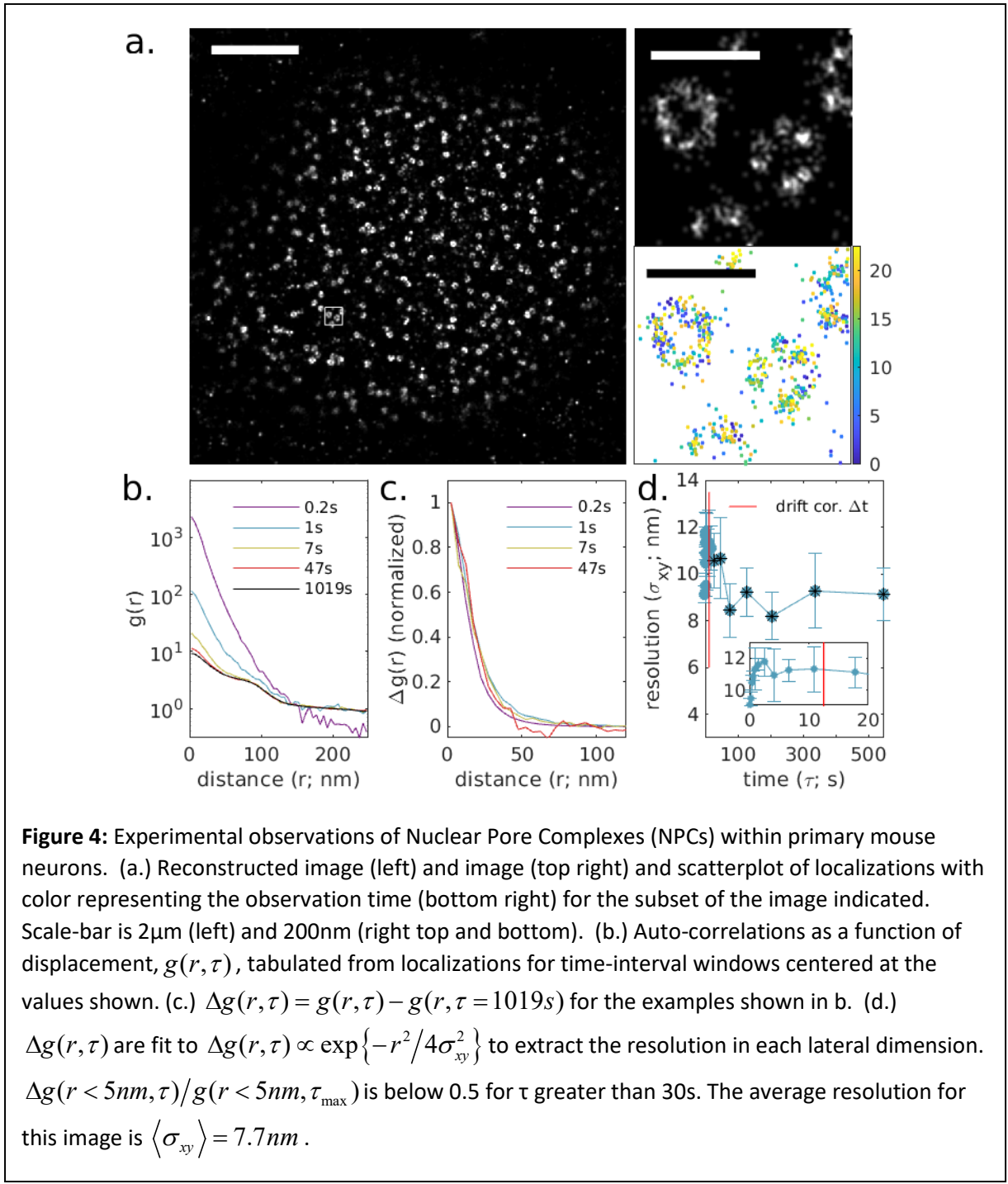

**Figure 4:** Experimental observations of Nuclear Pore Complexes (NPCs) within primary mouse neurons. (a.) Reconstructed image (left) and image (top right) and scatterplot of localizations with color representing the observation time (bottom right) for the subset of the image indicated. Scale-bar is 2μm (left) and 200nm (right top and bottom). (b.) Auto-correlations as a function of displacement, $g(r,\tau)$, tabulated from localizations for time-interval windows centered at the values shown. (c.) $\Delta g(r,\tau) = g(r,\tau) - g(r,\tau = 1019s)$ for the examples shown in b. (d.) $\Delta g(r,\tau)$ are fit to $\Delta g(r,\tau) \propto \exp\{-r^2/4\sigma_{xy}^2\}$ to extract the resolution in each lateral dimension. $\Delta g(r < 5nm, \tau)/g(r < 5nm, \tau_{max})$ is below 0.5 for τ greater than 30s. The average resolution for this image is $\langle \sigma_{xy} \rangle = 7.7nm$.

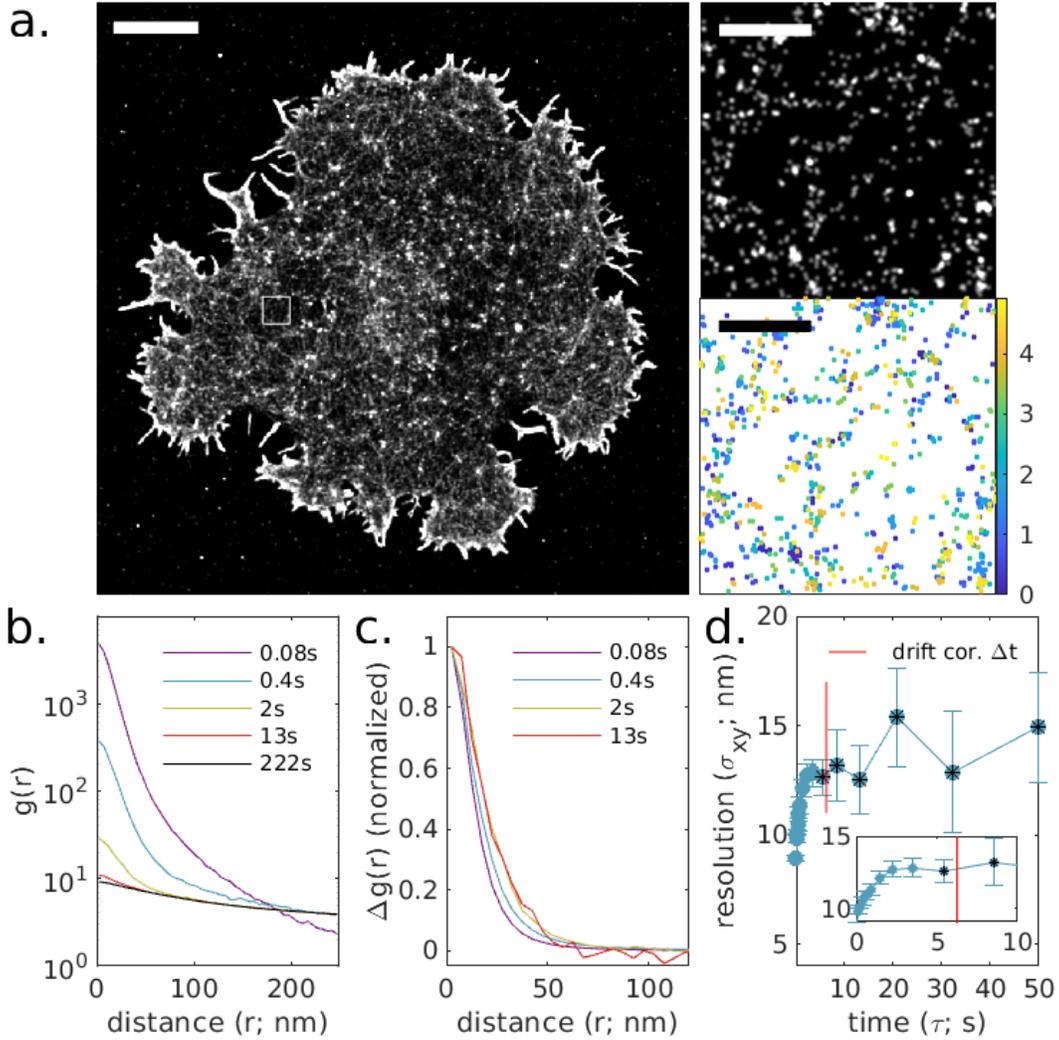

**Figure 5:** Experimental observations of f-actin on the ventral surface of a CH27 B cell. (a.) Reconstructed image (left) and image (top right) and scatterplot of localizations with color representing the observation time (bottom right) for the subset of the image indicated. Scale-bar is 5μm (left) and 500nm (right top and bottom). (b.) Auto-correlations as a function of displacement, $g(r,\tau)$, tabulated from localizations for time-interval windows centered at the values shown. (c.) $\Delta g(r,\tau) = g(r,\tau) - g(r,\tau = 222s)$ for the examples shown in b. (d.) $\Delta g(r,\tau)$ are fit to $\Delta g(r,\tau) \propto \exp\{-r^2/4\sigma_{xy}^2\}$ to extract the resolution in each lateral dimension. The time-scale of the drift correction is shown in red and $\Delta g(r<5nm,\tau)/g(r<5nm,\tau_{max})$ is below 0.5 for τ greater than 5.5s. The average resolution for this image is $\langle\sigma_{xy}\rangle = 13.5nm$.

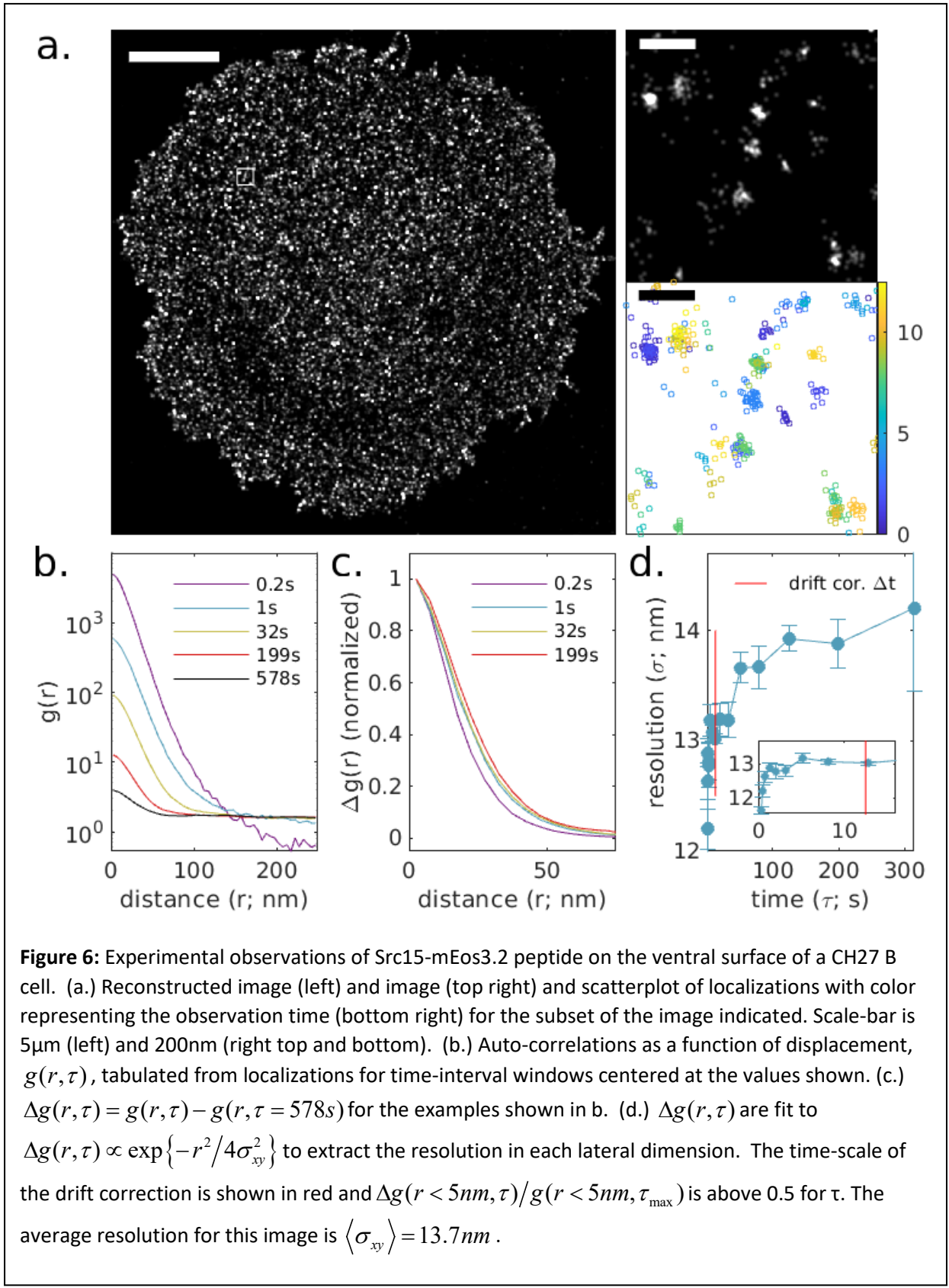

**Figure 6:** Experimental observations of Src15-mEos3.2 peptide on the ventral surface of a CH27 B cell. (a.) Reconstructed image (left) and image (top right) and scatterplot of localizations with color representing the observation time (bottom right) for the subset of the image indicated. Scale-bar is 5µm (left) and 200nm (right top and bottom). (b.) Auto-correlations as a function of displacement, $g(r,\tau)$, tabulated from localizations for time-interval windows centered at the values shown. (c.) $\Delta g(r,\tau) = g(r,\tau) - g(r,\tau = 578s)$ for the examples shown in b. (d.) $\Delta g(r,\tau)$ are fit to $\Delta g(r,\tau) \propto \exp\{-r^2/4\sigma_{xy}^2\}$ to extract the resolution in each lateral dimension. The time-scale of the drift correction is shown in red and $\Delta g(r<5nm,\tau)/g(r<5nm,\tau_{max})$ is above 0.5 for τ. The average resolution for this image is $\langle\sigma_{xy}\rangle = 13.7nm$.

Supplementary Material for:

# A lateral resolution metric for static single molecule localization microscopy images from time-resolved pair correlation functions

**Thomas R. Shaw[1,2], Frank J. Fazekas[1], Sumin Kim[3], Jennifer C. Flanagan-Natoli[1], Emily R. Sumrall[1], S. L. Veatch[1,2]**

[1]Program in Biophysics, [2]Program in Applied Physics, [3]Program in Cellular and Molecular Biology, University of Michigan, Ann Arbor, Michigan

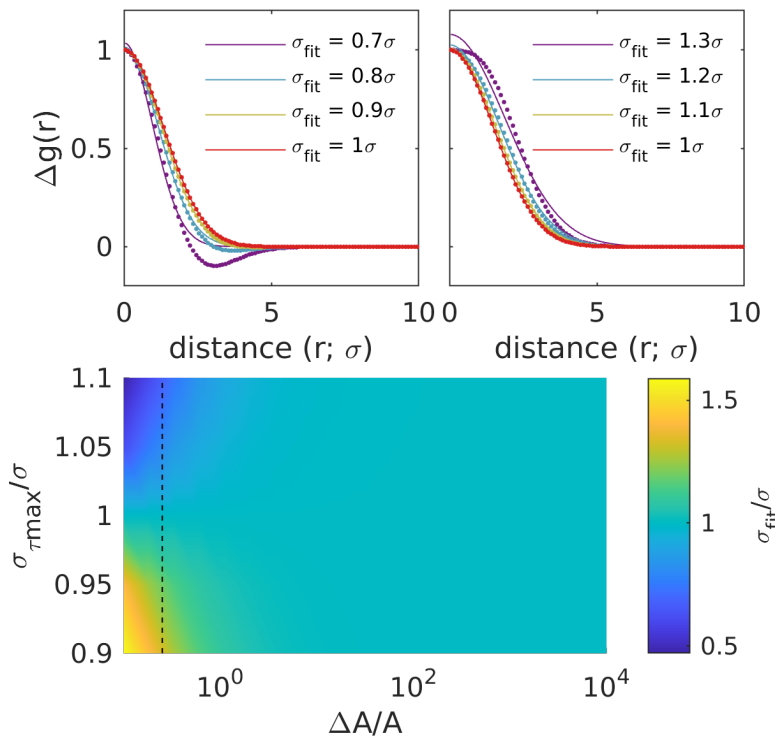

**Supplementary Figure S1:** Subtracting Gaussian shapes with different width leads to distortion in $\Delta g(r) = g(r,\tau) - g(r,\tau_{max})$ when $g(r,\tau)$ and $g(r,\tau_{max})$ have similar amplitudes but different widths. (top) plots of $\Delta g(r) = (A+\Delta A)\exp\{-r^2/4\sigma^2\} - A\exp\{-r^2/4\sigma_{\tau max}^2\}$ for $\sigma_{\tau max} = 1.1\sigma$ (left) and $\sigma_{\tau max} = 0.9\sigma$ (right) and $\Delta A = 0.25, 0.5, 1, 2$ from purple to red. Curves are normalized so they pass through 1 at r=0. The legend shows the width extracted when fitting $\Delta g(r)$ to a single Gaussian shape $\Delta g(r) = A\exp\{-r^2/4\sigma_{fit}^2\}$. A broader $\sigma_{\tau max}$ leads to systematic narrowing of $\sigma_{fit}$, while a narrow $\sigma_{\tau max}$ leads to systematic broadening of $\sigma_{fit}$ when the difference in amplitudes is order 1. (bottom) a summary of results over a broad range of $\Delta A$ and $\sigma_{\tau max}$ indicates that distortion is not a major concern over broad range of values interrogated.

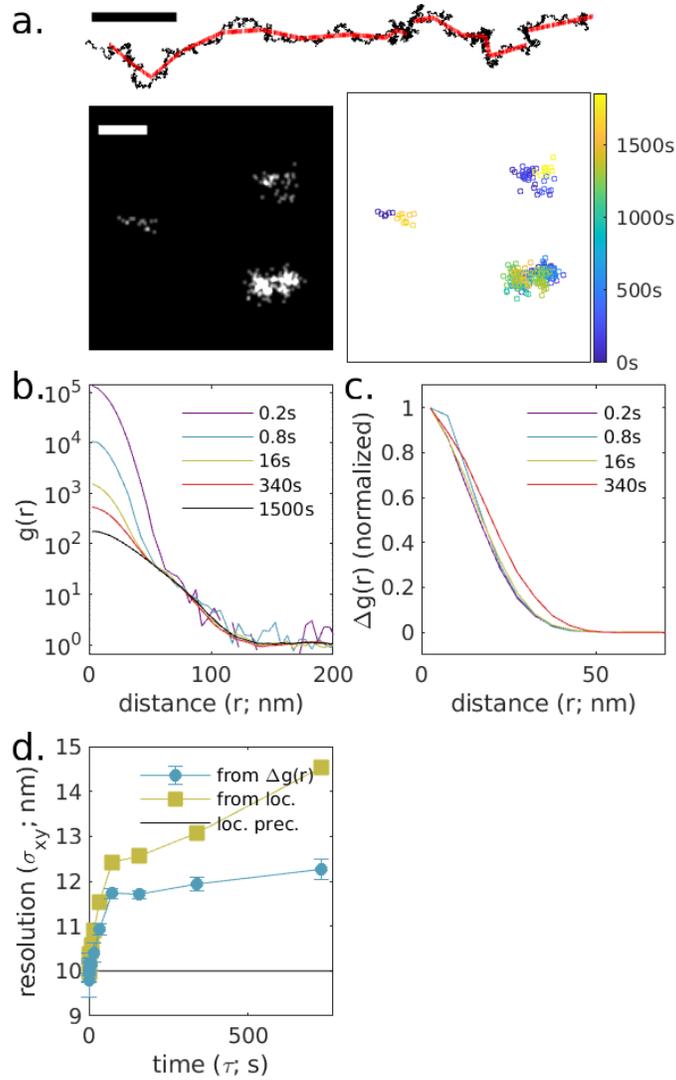

**Supplementary Figure S2: Simulation with drift and drift correction alongside single molecule motions.** (a.) The simulation from Fig 1 with applied drift (black) and drift correction (red) as shown in the trajectory above as well as single molecule diffusion with D=1nm$^2$/sec. Reconstructed image (left) and scatterplot of localizations with color representing the observation time (right) for a small subset of the simulated plane. Scale-bar is 100nm. (b.) Auto-correlations as a function of displacement, $g(r,\tau)$, tabulated from simulations for time-interval windows centered at the values shown. (c.) $\Delta g(r,\tau) = g(r,\tau) - g(r,\tau = 1500s)$ for the examples shown in b. (d.) $\Delta g(r,\tau)$ are fit to $\Delta g(r,\tau) \propto \exp\{-r^2/4\sigma_{xy}^2\}$ to extract out the resolution in each lateral dimension (from $\Delta g(r)$). The resolution estimated from $\Delta g(r,\tau)$ varies with time-interval, and is systematically narrower than the point spread function measured by grouping localizations with molecules ((from loc.). This is due to the distortion effect demonstrated in Fig S1 and is characterized by a $\sigma_{xy}$ that increases with $\tau$.

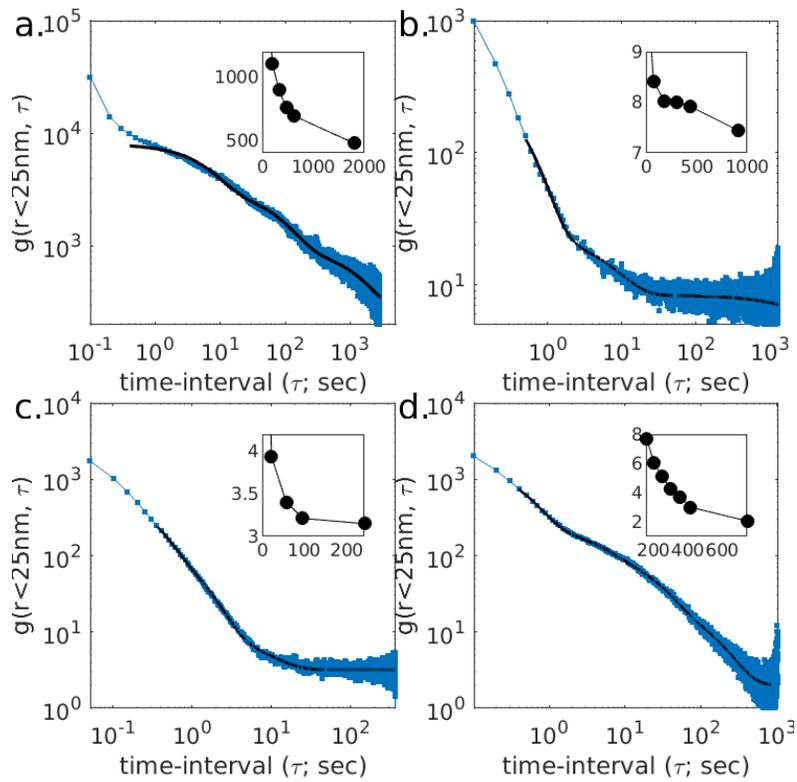

**Supplementary Figure 3.** Plots of $g(r<50nm,\tau)$ for the experimental samples shown in Figs 3-6: (a.) DNA origami rulers from Fig 3, (b.) Alexa647 labeled NUP 210 from Fig 4, (c.) Alexa647-phalloidin from Fig 5. (d.) mEeos3.2 conjugated Src15 from Fig 6. These curves capture $g_e(\tau)$ up to a numerical offset that is dependent on the structure present in the image. Black lines are fit to a sum of exponentials and are present to highlight the monotonically decreasing trend.

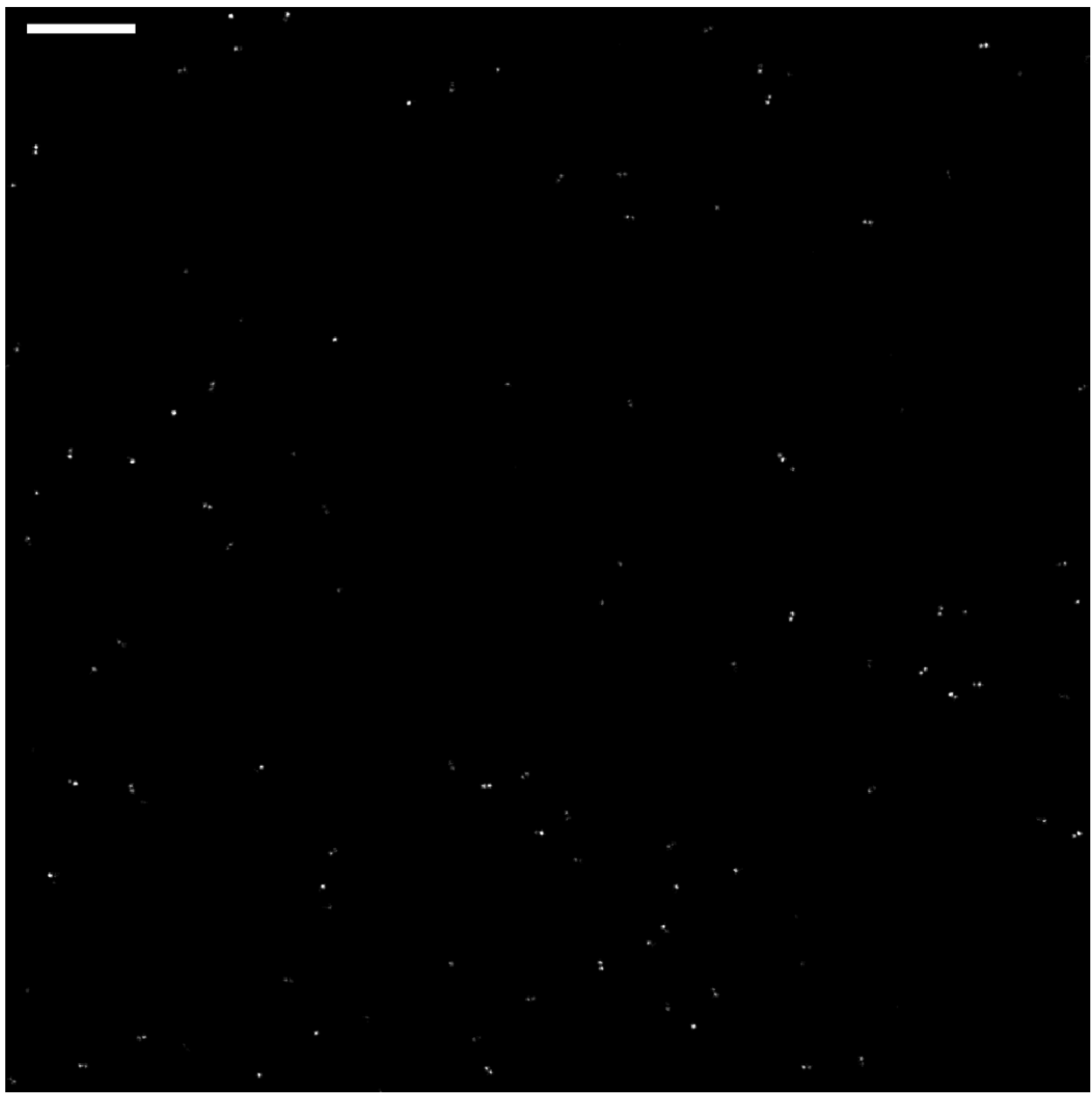

**Supplementary Figure S4.** 10μm by 10μm region showing simulated localizations from Figs 1-2. The full simulated area was 40μm by 40μm. Scale bar is 1μm.

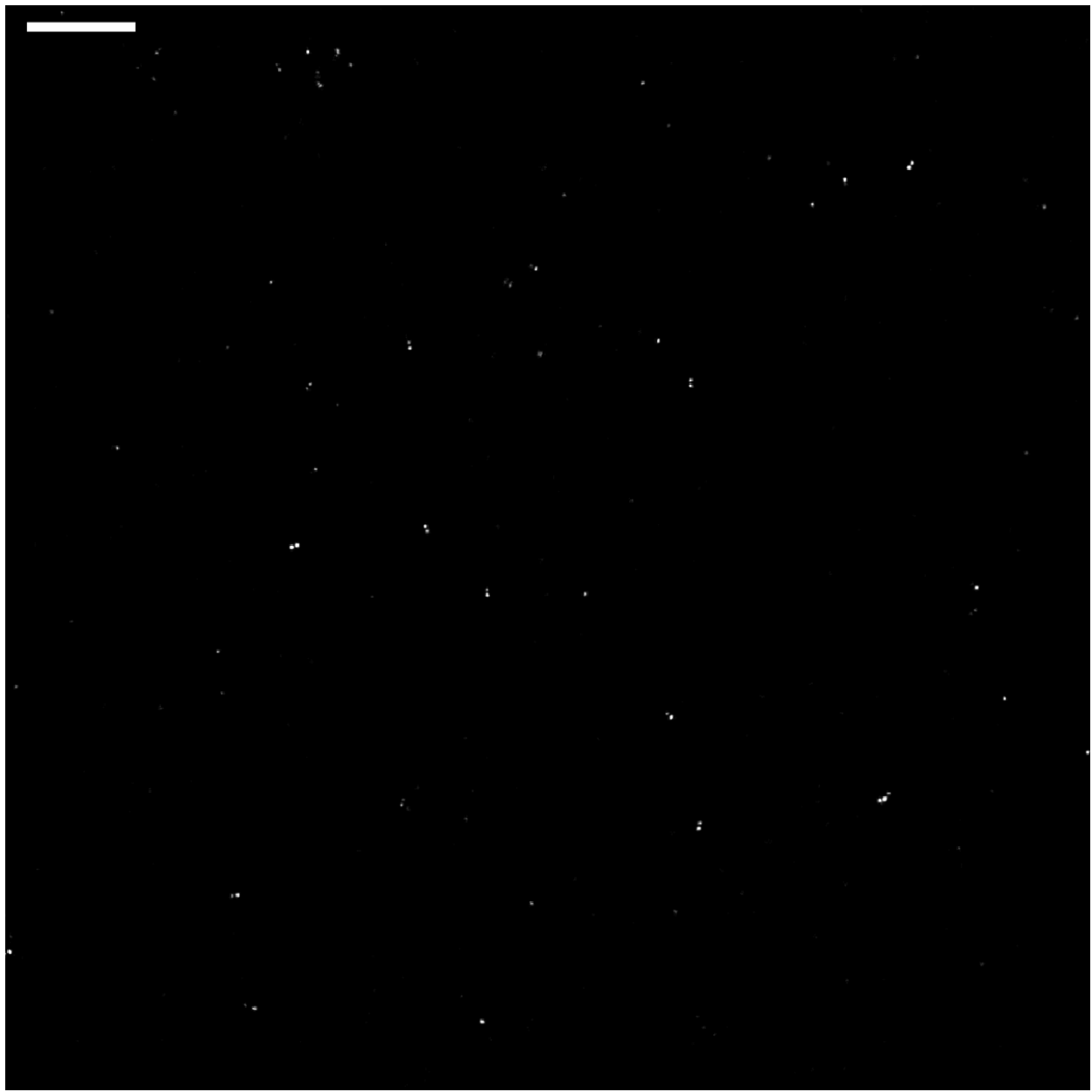

**Supplementary Figure S5.** 10μm by 10μm region showing DNA origami rulers analyzed in Fig 3. The full imaged area was 40μm by 40μm. Scale bar is 1μm.

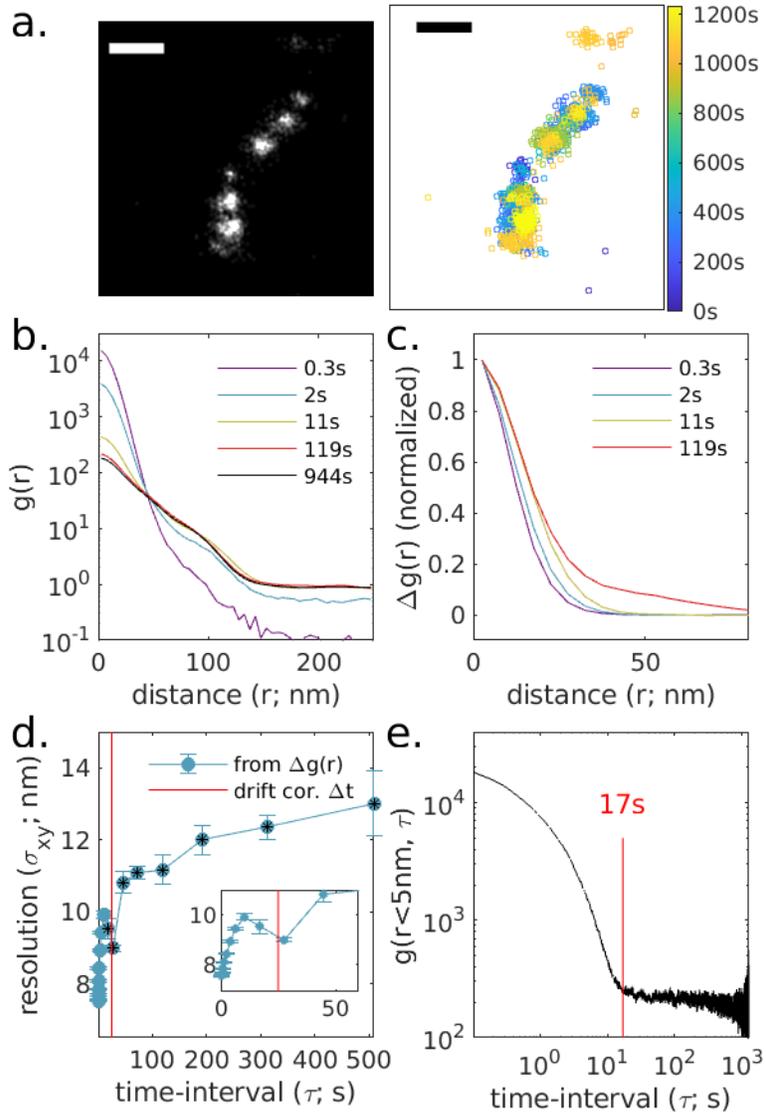

**Supplementary Figure S6: Analysis of DNA origami imaged using DNA PAINT.** Experimental observations of DNA origami rulers made up of 3 linear docking strands separated by 40nm. (a.) Reconstructed image (left) and scatterplot of localizations with color representing the observation time (right) for a small subset of the observed plane. Scale-bar is 100nm and 50nm in the inset. (b.) Auto-correlations as a function of displacement, $g(r,\tau)$, tabulated from localizations for time-interval windows centered at the values shown. (c.) $\Delta g(r,\tau) = g(r,\tau) - g(r,\tau = 944s)$ for the examples shown in b. (d.) $\Delta g(r,\tau)$ are fit to $\Delta g(r,\tau) \propto \exp\{-r^2/4\sigma_{xy}^2\}$ to extract the resolution in each lateral dimension (from $\Delta g(r)$) which varies with time-interval. $\Delta g(r < 5nm, \tau)/g(r < 5nm, \tau_{max})$ falls below 0.5 for τ greater than 17s. The gradual increase in $\sigma_{xy}$ is an indication that the approximation of Eqn. 4 is no longer valid, since resolution is not expected to vary due to drift at these long time-scales. (e.) Temporal correlations become negligible near τ=17s in this sample.

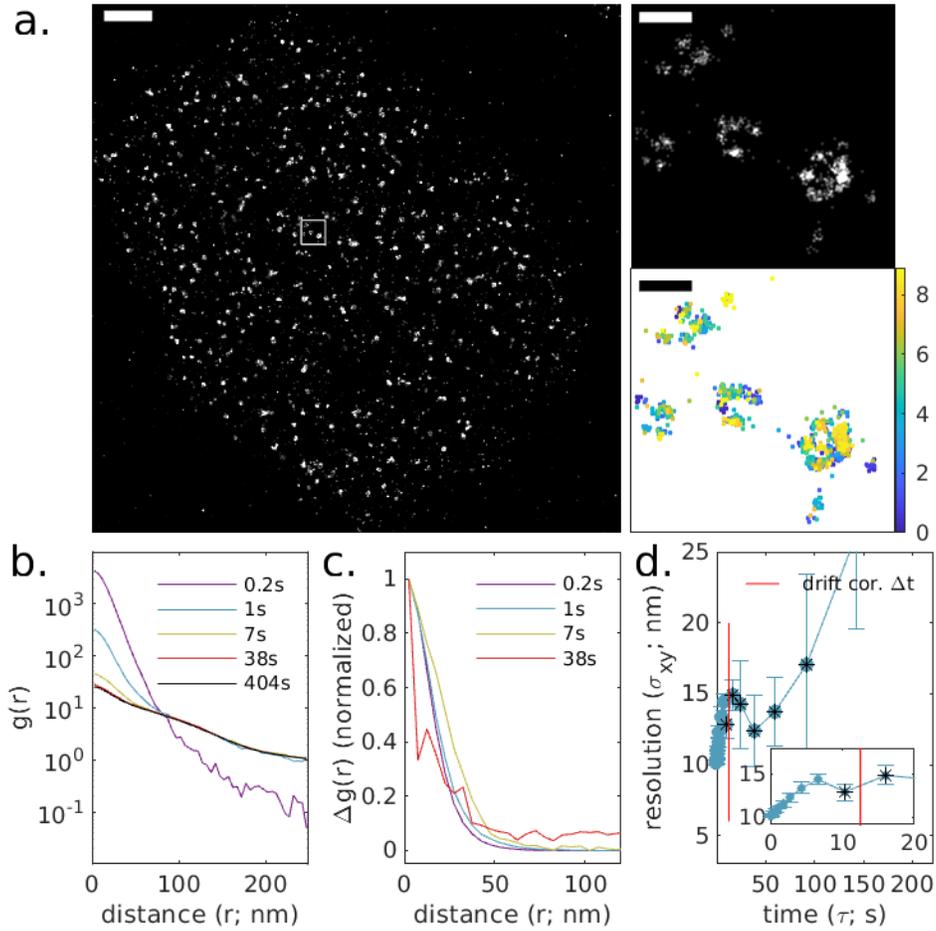

**Supplementary Figure S7:** Experimental observations of Nuclear Pore Complexes clathrin coated pits in CH27 B cells imaged using DNA PAINT. (a.) Reconstructed image (left) and image (top right) and scatterplot of localizations with color representing the observation time (bottom right) for the subset of the image indicated. Scale-bar is 2μm (left) and 200nm (right top and bottom). (b.) Auto-correlations as a function of displacement, $g(r,\tau)$, tabulated from localizations for time-interval windows centered at the values shown. (c.) $\Delta g(r,\tau) = g(r,\tau) - g(r,\tau = 404s)$ for the examples shown in b. (d.) $\Delta g(r,\tau)$ are fit to $\Delta g(r,\tau) \propto \exp\{-r^2/4\sigma_{xy}^2\}$ to extract the resolution in each lateral dimension. $\Delta g(r<5nm,\tau)/g(r<5nm,\tau_{max})$ is below 0.5 for τ greater than 10s. Beyond $\tau > 50s$, the estimate of $\sigma_{xy}$ begins increasing steadily, consistent with failure of $\Delta g(r,\tau)$ to provide a reasonable of $g_{PSF}(r,\tau)$.

**Supplementary Note: Derivation of spacetime pair correlation function estimator, and related computations.**

$N$ localizations $\mathbf{u}_i = (\vec{r}_i, t_i) = (x_i, y_i, t_i)$, $i = 1, \ldots, N$ are observed on a spatial window (region of interest/ROI) $W$ during a temporal window $T$. This set of points is considered as a realization of a space-time point process $X$, so that we may define a (first-order) density $\rho(\mathbf{u}) = \rho(\vec{r}, t)$ notionally as

$$\frac{\text{Expected \# of points in area } d\vec{r} \text{ and time-interval } dt \text{ around } (\vec{r}, t)}{d\vec{r} \cdot dt}$$

Or more formally by the following:

$$\mathrm{E} \sum_{\mathbf{u} \in X \cap W \times T} 1[\mathbf{u} \in A] = \int_A \rho(\mathbf{u}) d\mathbf{u} \qquad (0.1)$$

For any set $A \subset W \times T$, where $1[\cdot]$ is an indicator function, taking the value 1 when its argument is true, and 0 otherwise. For the purposes of this paper, we assume that $\rho = \rho(t)$ is constant in space but may vary in time, e.g. due to bleaching of the fluorophores of the sample.

Further, define the second-order density $\rho^{(2)}(\mathbf{u}_1, \mathbf{u}_2)$ notionally by

$$\frac{\text{Expected \# of pairs of points in the } d\vec{r} \cdot dt \text{ neighborhoods of } \mathbf{u}_1 \text{ and } \mathbf{u}_2 \text{ respectively}}{(d\vec{r} \cdot dt)^2}$$

Or more formally

$$\mathrm{E} \sum_{\mathbf{u}_1, \mathbf{u}_2 \in X \cap W \times T}^{\neq} 1[\mathbf{u}_1 \in A \text{ and } \mathbf{u}_2 \in B] = \int_A \int_B \rho^{(2)}(\mathbf{u}_1, \mathbf{u}_2) d\mathbf{u}_2 d\mathbf{u}_1 \qquad (0.2)$$

Now $\rho^{(2)}$ describes the second-order properties of $X$, for example attraction or repulsion between points. It is convenient to normalize $\rho^{(2)}$ so that it is dimensionless and easier to interpret. To that end, define the pair autocorrelation function $g(\mathbf{u}_1, \mathbf{u}_2)$:

$$g(\mathbf{u}_1, \mathbf{u}_2) = \frac{\rho^{(2)}(\mathbf{u}_1, \mathbf{u}_2)}{\rho(\mathbf{u}_1, \mathbf{u}_2)} \qquad (0.3)$$

Loosely, the pair autocorrelation function is the ratio of the actual probability of finding points at both $\mathbf{u}_1$ and $\mathbf{u}_2$ to the hypothetical probability under the assumption that $\mathbf{u}_1$ and $\mathbf{u}_2$ are independent. We typically assume that $g$ is translation invariant in both space and time, and often further assume that it is rotationally invariant in space, so that it only depends on the separation of u1 and u2 in space and time, and we may write $g(\mathbf{u}_1, \mathbf{u}_2) = g(\|\vec{r}_2 - \vec{r}_1\|, t_2 - t_1)$.

We estimate $g$ using the standard kernel-based framework as laid out in e.g. (1, 2). Specifically, we use a box kernel with bandwidth $\delta_r$ in space and $\delta_t$ in time, and an isotropic edge-correction in space, and a density correction for the temporal edge correction, following the approach of (3). Briefly, consider the family of estimators for $g(r,\tau)$ given by:

$$\hat{g}(r,\tau) := \frac{1}{\gamma_{sp.}(r)\gamma_t(\tau)} \sum_{\mathbf{u}_i, \mathbf{u}_j \in X \cap W \times T}^{\neq} 1[|\|\vec{r}_j - \vec{r}_i\| - r| < \delta_r/2, |t_j - t_i - \tau| < \delta_t/2] \qquad (0.4)$$

We wish to derive functions $\gamma_{sp.}$ and $\gamma_t$ such that the resulting estimator is unbiased. The expectation value of the sum in the above expression can be determined from an appropriate Campbell's theorem:

$$\mathrm{E}\hat{g}(r,\tau) = \frac{1}{\gamma_{sp.}\gamma_t} \int_{W\times T}\int_{W\times T} \rho^{(2)}(\mathbf{u}_1, \mathbf{u}_2) 1[|\|\vec{r}_2 - \vec{r}_1\| - r| < \delta_r/2, |t_2 - t_1 - \tau| < \delta_t/2] d\mathbf{u}_1 d\mathbf{u}_2$$

$$= \frac{1}{\gamma_{sp.}\gamma_t} \int_{W\times T}\int_{W\times T} g(\|\vec{r}_2 - \vec{r}_1\|, t_2 - t_1)\rho(t_1)\rho(t_2) 1[|\|\vec{r}_2 - \vec{r}_1\| - r| < \delta_r/2] 1[|t_2 - t_1 - \tau| < \delta_t/2] d\mathbf{u}_1 d\mathbf{u}_2$$

$$\approx \frac{g(r,\tau)}{\gamma_{sp.}\gamma_t} \int_W\int_W 1[|\|\vec{r}_2 - \vec{r}_1\| - r| < \delta_r/2] d\vec{r}_1 d\vec{r}_2 \int_T\int_T \rho(t_2)\rho(t_1) 1[|t_2 - t_1 - \tau| < \delta_t/2] dt_1 dt_2$$

Where the approximation in line 3 is due to the assumption that $g(r,\tau)$ is almost constant within $\delta_r/2$ in space and $\delta_t/2$ in time.

From the above derivation, it follows that $\hat{g}(r,\tau)$ is unbiased for the choices

$$\gamma_{sp.}(r) = \int_W\int_W 1[|\|\vec{r}_2 - \vec{r}_1\| - r| < \delta_r/2] d\vec{r}_2 d\vec{r}_1$$

$$\gamma_t(\tau) = \int_T\int_T \rho(t_1)\rho(t_2) 1[|t_2 - t_1 - \tau| < \delta_t/2] dt_2 dt_1$$

For computational considerations, we make further approximations on $\gamma_{sp.}$:

$$\gamma_{sp.}(r) = \int_W \int 1[|\|\vec{h}\| - r| < \delta_r/2] 1[\vec{r}_1 + \vec{h} \in W] d\vec{h} d\vec{r}_1$$

$$= \int_W \int_0^{2\pi} \int 1[|h - r| < \delta_r/2] 1[\vec{r}_1 + (h\cos\theta, h\sin\theta) \in W] h\, dh\, d\theta\, d\vec{r}_1$$

$$= \int_W \int_{r-\delta_r/2}^{r+\delta_r/2} h \int_0^{2\pi} 1[\vec{r}_1 + (h\cos\theta, h\sin\theta) \in W] d\theta\, dh\, d\vec{r}_1$$

$$\approx r\delta_r \int_W \int_0^{2\pi} 1[\vec{r}_1 + (r\cos\theta, r\sin\theta) \in W] d\theta\, d\vec{r}_1$$

$$= r\delta_r \int_0^{2\pi} |W \cap W_{-(r\cos\theta, r\sin\theta)}| d\theta$$

where $|A|$ indicates the area of the set $A$, and $A_{\vec{h}}$ indicates the translation of the set $A$ by the vector $\vec{h}$. The first line is a change of variables to $\vec{h} = \vec{r}_2 - \vec{r}_1$, with the extra indicator functions reflecting the integration bounds on $\vec{r}_2$, followed by a change to polar coordinates for $\vec{h}$. The approximation in the fourth line is justified by the fact that angular integral varies slowly with $h$, so the radial part of the integral can be approximately separated.

For the purposes of our matlab code, we represent the spatial window/ROI $W$ as a polygon with vertices [mask.x(i),mask.y(i)]. We translate the ROI by a vector [hx,hy] by simply adding hx and hy to mask.x and mask.y, respectively. Matlab provides functions polybool (to compute the intersection $W \cap W_{-\vec{h}}$, and polyarea to compute the area of the resulting polygon. It remains to complete the angular integral, which we compute by discretizing theta into 32 equally spaced points

$$\gamma_{sp.}(r) = \frac{2\pi r \delta_r}{32} \sum_{i=1}^{32} |W \cap W_{-(r\cos\theta_i, r\sin\theta_i)}|, \qquad \theta_i = \frac{2\pi i}{32}$$